\documentstyle[12pt,aaspp4,psfig]{article}
\def\ltsima{$\; \buildrel < \over \sim \;$}
\def\simlt{\lower.5ex\hbox{\ltsima}}
\def\gtsima{$\; \buildrel > \over \sim \;$}
\def\simgt{\lower.5ex\hbox{\gtsima}}
\def\arcsec{\mathop{\rm arcsec}\nolimits} 

\def\kms{{\rm\,km\,s^{-1}}}

\def\msun{{\rm\,M_\odot}}
\def\lsun{{\rm\,L_\odot}}

\def\s{\ifmmode \widetilde \else \~\fi}
\def\={\overline}

\def\spose#1{\hbox to 0pt{#1\hss}}

\def\etal{{\it et al.\ }}

\def\lta{\mathrel{\spose{\lower 3pt\hbox{$\mathchar"218$}}
     \raise 2.0pt\hbox{$\mathchar"13C$}}}
\def\gta{\mathrel{\spose{\lower 3pt\hbox{$\mathchar"218$}}
     \raise 2.0pt\hbox{$\mathchar"13E$}}}
\def\Dt{\spose{\raise 1.5ex\hbox{\hskip3pt$\mathchar"201$}}}    
\def\dt{\spose{\raise 1.0ex\hbox{\hskip2pt$\mathchar"201$}}}    

\def\=={\equiv}

\def\dotsfill{\leaders\hbox to 1em{\hss.\hss}\hfill}

\slugcomment{Submitted to the Astronomical Journal}

\newcommand{\ffffff}[1]{\mbox{$#1$}}
\newcommand{\scnd}{\mbox{\ffffff{''}\hskip-0.3em .}}

\newcommand{\apm}{APM~08279+5255}

\newcommand{\lya}{Ly$_\alpha$}

\lefthead{Ibata et al.}
\righthead{\apm}

\begin{document}

\title{NICMOS and VLA Observations of the Gravitatonally Lensed 
Ultraluminous BAL Quasar \apm:\\
Detection of a Third Image}

\author{
Rodrigo A. Ibata\altaffilmark{1},
Geraint F. Lewis\altaffilmark{2},
Michael J. Irwin\altaffilmark{3},
Joseph Leh\'{a}r\altaffilmark{4},
Edward J. Totten\altaffilmark{5}}
 
\altaffiltext{1}
{European Southern Observatory, Garching bei M\"unchen, Germany \\
Electronic mail: {\tt ribata@eso.org}}

\altaffiltext{2}{ 
Fellow of the Pacific Institute for Mathematical Sciences 1998-1999, \\
Dept. of Physics and Astronomy, University of Victoria, Victoria, B.C., Canada 
\& \\ 
Astronomy Dept., University of Washington, Seattle WA, U.S.A.
\\
Electronic mail: {\tt gfl@uvastro.phys.uvic.ca} \\ 
Electronic mail: {\tt gfl@astro.washington.edu}}

\altaffiltext{3}
{Institute of Astronomy, Madingley Road, Cambridge, CB3 0HA, U.K. \\
Electronic mail: {\tt mike@ast.cam.ac.uk}}
 
\altaffiltext{4}
{Havard-Smithsonian Center for Astrophysics, Cambridge, MA, U.S.A. \\
Electronic mail: {\tt jlehar@cfa.harvard.edu}}
 
\altaffiltext{5}
{Department of Physics, Keele University, Keele,
Staffordshire, UK \\
Electronic mail: {\tt ejt@astro.keele.ac.uk}}
 
\begin{abstract}
We  present  a  suite  of  observations  of  the  recently  identified
ultraluminous BAL  quasar \apm, taken  both in the infra-red  with the
NICMOS high resolution camera on board the Hubble Space Telescope, and
at 3.5cm  with the Very Large  Array.  With an  inferred luminosity of
${\rm \sim5\times10^{15}\lsun}$, \apm\ is apparently the most luminous
system  known.  Extant  ground-based  images show  that  \apm\ is  not
point-like, but  is instead separated into  two components, indicative
of gravitational lensing.  The much higher resolution images presented
here  also  reveal  two  point  sources,  A and  B,  of  almost  equal
brightness  ($f_{\rm  B}/f_{\rm  A}=0.782  \pm 0.010$),  separated  by
$0\scnd378  \pm 0\scnd001$, as  well as  a third,  previously unknown,
fainter image, C, seen between  the brighter images.  While the nature
of C  is not fully determined,  several lines of evidence  point to it
being a third gravitationally lensed  image of the quasar, rather than
being the  lensing galaxy.  Simple models which  recover the relative
image configuration and brightnesses  are presented.  While proving to
be  substantially amplified, \apm\  possesses an  intrinsic bolometric
luminosity of $\sim10^{14}\rightarrow10^{15}\lsun$ and remains amongst
the most luminous objects known.
\end{abstract}


\section{Introduction}\label{introduction}
Discovered  serendipitously in  a survey  of carbon  stars  within the
Galactic halo, the bright ${\rm (m_r = 15.2)}$ z=3.87 broad absorption
line  quasar \apm\  was found  to  be positionally  coincident with  a
source in the IRAS Faint Source Catalog (\cite{ir98}).  The bolometric
luminosity,  inferred from  these optical  and far-IR  fluxes, exceeds
${\rm   5\times10^{15}\lsun  ~(\Omega_o=1,  H_o=50~km~s^{-1}~Mpc^{-1},
{\rm  assumed throughout})}$,  making \apm\  the most  luminous object
currently known.  Submillimeter  photometry is consistent with thermal
emission  from a  massive ${\rm  (\sim 3\times10^8\msun)}$,  warm dust
component  (\cite{le98}), with  \apm\ displaying  an  overall spectral
energy distribution which is similar to other ultraluminous systems at
high redshift (e.g.  H1413+117~\cite{ba95}).  Recent observations with
the  30m IRAM  telescope  have  also detected  a  massive quantity  of
molecular gas;  such an  environment is ripe  for star  formation, and
this process may  indeed be responsible for a  substantial fraction of
\apm's   phenomenal   luminosity   (\cite{do98}).    From   their   CO
observations, Downes et al.\ (1998) determined a redshift of 3.911 for
\apm.  This  differs by $\sim2500\kms$  from that determined  from the
high  ionization  broad emission  lines  (\cite{ir98}).   Since it  is
essentially impossible to measure  accurately the systemic redshift of
a complex BAL  QSO like \apm\ from optical spectra,  we have chosen to
adopt the value of z=3.911 throughout this work.

Gravitational lensing  can distort our  view of the  distant universe,
enhancing the flux of galaxies  and AGN, giving them the appearance of
extraordinary   systems.   This   has  been   the  case   for  several
ultraluminous  systems  discovered in  recent  years [e.g.   H1413+117
(\cite{ma88,kn98}),  IRAS FSC  10214+4724  (\cite{ro91,br95,ei96}) and
the proto-galaxy candidate MS1512-cB58 (\cite{ye96,wi96,se98})]. Given
the  extreme  inferred  luminosity   of  \apm,  the  possibility  that
gravitational lensing is influencing the observed properties must also
be considered.

Analysis of  the point  spread function (PSF)  derived from  the first
ground-based observations, obtained with the Jacobus Kapteyn Telescope
(JKT), revealed  that \apm\ possesses a  non-stellar profile.  Rather,
its profile was found to be better represented by a pair of point-like
sources,   separated  by   $\sim0\scnd4$   (\cite{ir98}).   This   was
interpreted as  indicative of multiple  imaging of \apm\ by  a massive
system along  the line-of-sight,  probably one or  both of  the strong
\ion{Mg}{2} absorbers  seen at z=1.18 and z=1.81.   The two components
were  found to possess  very similar  brightnesses, within  10\%; with
this, the magnification  of the optical continuum was  estimated to be
$\sim40$.  Follow-up observations with the adaptive optics bonnette at
the Canada-France-Hawaii telescope confirmed the existence of multiple
components  in  \apm,   with  a  separation  of  $0\scnd35\pm0\scnd02$
(\cite{ld98}).  The relative brightness of the two images was found to
be   $1.21\pm0.25$,  consistent  with   the  JKT   observations.   The
submillimeter-infrared flux, which arises in a larger emission region,
is subject  to less enhancement  [for IRAS FSC 10214+4724  the optical
flux is thought to be magnified a factor of $2\rightarrow3$ times that
of the dominant infrared emission (\cite{ei96})].

While the above  observations revealed that \apm\ is  a good candidate
for  gravitational lensing,  the scale  of the  PSF in  the  images is
comparable to  the image separation,  and no detection of  the lensing
galaxy  was   made.   The  uncertainty  of   the  relative  image/lens
configuration leads  to uncertainty  in the lens  model and  hence the
inferred lensing magnification, without which the intrinsic properties
of \apm\  cannot be determined.  In  an effort to  confirm the lensing
hypothesis and identify the lensing lensing galaxy, \apm\ was observed
with NICMOS  on the Hubble Space  Telescope and with the  VLA, and the
results    of   these   observations    are   presented    here.    In
Section~\ref{observations}  the   details  of  the   observations  are
presented,   while  a   lens   model  for   \apm\   is  presented   in
Section~\ref{lens}.  The  conclusions of  this study are  presented in
Section~\ref{conclusions}.

\section{Observations}\label{observations}

\subsection{NICMOS}
On   the  11th   of  October   1998,  we   observed  \apm\   with  the
NICMOS\footnote{ Based on observations  with the NASA/ESA Hubble Space
Telescope, obtained at the Space Telescope Science Institute, which is
operated by the Association of Universities for Research in Astronomy,
Inc. under  NASA contract No. NAS5-26555.}  infra-red  camera on board
the  Hubble Space  Telescope  (HST).  The  highest spatial  resolution
NICMOS camera NIC1  was used to obtain 12 exposures  of 14~sec each in
the  J-band (F110W), and  12 exposures  of 40~sec  each in  the H-band
(F160W).  A  further 12 exposures of  10~sec each were  secured in the
K-band  (F205W)  with the  NIC2  camera (as  a  K-band  filter is  not
available on NIC1).  Each set  of exposures were taken with 1'' offset
pointings in  a spiral dither  pattern.  The main difference,  for the
present  purposes, between  the NIC1  and  NIC2 cameras  is the  pixel
scale: the pixel scale is $0\scnd043$ on NIC1 and $0\scnd075$ on NIC2.
The orientation of the instrument  was such that the position angle of
the CCD y-axis on all exposures with the NIC1 camera was $66.84^\circ$
E of  N, and all  exposures with the  NIC2 camera have the  CCD y-axis
$66.03^\circ$ E of N.

The raw  frames were pre-reduced  with the package CALNICA,  using the
appropriate calibration files from the STScI archive.  This processing
includes  bias  subtraction, flat  fielding,  dark  correction, and  a
photometric  calibration for  converting detector  counts  to physical
flux units.

\begin{figure*}
\centerline{
\psfig{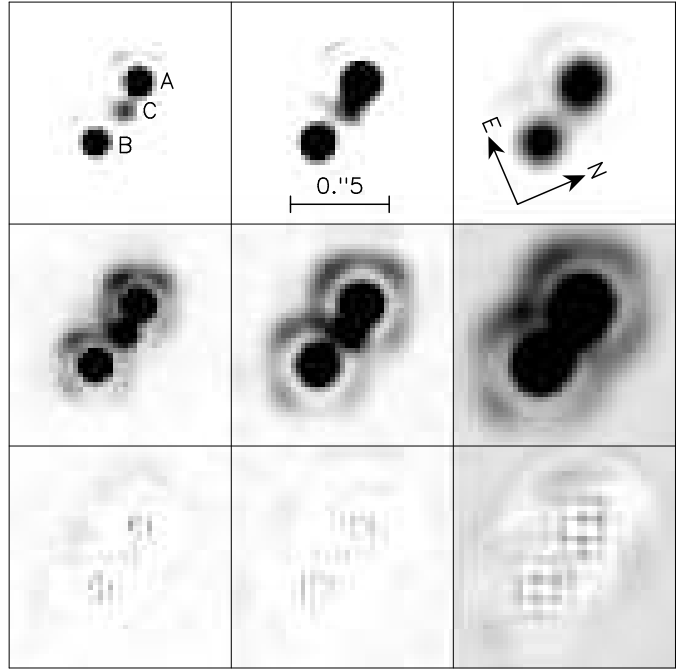}
}
\caption[]{The NICMOS images of \apm.  From left to right, the columns
show the oversampled  combined images in the filters  F110W, F160W and
F205W. The same  $1\scnd125 \times 1\scnd125$ region is  shown in each
panel.  In the top-row panels, the brightness cuts have been chosen to
show the  third image, which  is clearly visible between  the brighter
two images.   The middle-row panels have brightness  cuts to emphasize
the first  Airy rings. The residuals  of the PSF  fitting (detailed in
the text) are shown in the bottom-row panels.}
\label{fig1}
\end{figure*}

A visual  inspection of  the preprocessed frames  (see the top  row of
Figure~\ref{fig1})   revealed   two   bright   images   separated   by
$0\scnd35$--$0\scnd4$,   as  previously  inferred   from  ground-based
images.  The brightest image, which we hereafter refer to as component
``A'', is located towards the NE; the slightly fainter component ``B''
is on the  SW end of the system.  Clearly  visible on the preprocessed
F110W and F160W frames is a third, previously undetected, image.  This
component (which we shall  label ``C''), is substantially fainter than
components A and B, and is located between those two bright images.

The gravitational  lensing models  which are compared  to the  data in
\S\ref{lens} below require an accurate description of the observations
in terms  of relative fluxes and  positions of the  components of this
system. To this  end we have analyzed the data  in two alternate ways,
measuring magnitudes and  positions on individual preprocessed frames,
and on a combined image stack in each passband.

Since the three components are  at best $\sim 3$ pixels separated from
each other, it is necessary to perform a careful point-spread function
fitting analysis,  a requirement  of which is  the determination  of a
good PSF model for each camera and passband. The NICMOS PSF depends on
many  factors.   The  factors  that  arise  from  spacecraft  settings
include:  the chosen  camera and  filter, the  focus setting,  and the
position  of  the source  on  the  detector.   The PSF  is  wavelength
dependent, so  the source spectrum also  affects its PSF.   It was for
this  last reason  that we  adopted  PSFs simulated  with the  TINYTIM
algorithm (which models the effects  of all of the above constraints),
instead of picking stars from the STScI archive of NICMOS observations
to construct a PSF.  We assumed that \apm\ has a flat spectrum in $\nu
F_\nu$ over  the region from  $0.8 \mu m  \rightarrow 2.35 \mu  m$, as
suggested  by  the spectral  energy  distribution  displayed in  Lewis
\etal\  (1998),  their Figure~2.   This  choice  is  supported by  the
photometric results of the present study, listed in Table~1 below.

In the  first pass  of data-reductions, we  constructed a  TINYTIM PSF
appropriate for the camera,  filter, focus position and assumed source
spectrum, and  used the  approximate location of  component A  in each
frame as the PSF  position. This PSF was then fit to  the data on each
data frame  using the  ALLSTAR PSF fitting  program (Stetson  1987) to
measure the  magnitudes and positions of the  three components.  These
position   measurements  are   accurate  to   typically   better  than
$0.1$~pixels     (judging     from      the     RMS     scatter     in
$|\vec{x_A}-\vec{x_B}|$).  This information  enables us to use TINYTIM
to  construct better PSFs,  taking into  account the  actual sub-pixel
location  of each  component on  each frame.   Refined  magnitudes and
positions  were  subsequently   obtained  by  re-running  the  ALLSTAR
program; the mean  and RMS values of these  measurements are listed in
Table~1.

A  combined stacked  frame  was constructed,  using  the positions  of
components A  and B  on each frame  to define the  frame registration.
The individual  preprocessed data frames  were resampled onto  a finer
grid  at  a  scale   of  $0\scnd025/{\rm  pixel}$,  using  the  STSDAS
``DRIZZLE'' algorithm, and  then medianed to give a  combined frame in
each of  the F110W, F160W  and F205W passbands; these  high resolution
frames  are  reproduced  in  Figure~\ref{fig1},  and are  shown  as  a
color-composite map in Figure~2.

In   the  middle-row   panels  of  Figure~\ref{fig1}  we  have  chosen image
brightness cuts that emphasize a peculiarity of  our dataset: the first Airy
ring is not uniform,  but instead appears brighter  on the higher row number
side of the object centers. This feature of the PSF is  not predicted by the
TINYTIM  software\footnote{The anonymous referee   brought to  our attention
that  other  NICMOS data (e.g. the   F160W NIC2 exposures of RXJ0911.4+0551)
also display a non-uniform Airy ring.}.
Note that it  cannot be due to some structure of
the  source,  as  it  is   coincident  with  the  Airy  ring  in  each
passband. The deviation  away from the model PSF  is significant --- a
normal TINYTIM  PSF model  that fits  the core of  component A  in the
F110W  filter  underestimates the  flux  in  the  first Airy  ring  by
approximately 10\%  of the  total flux (i.e.   the first Airy  ring is
approximately 50\%  brighter than expected).  We  implemented a simple
fix of  this problem by altering  the model PSFs: we  applied a linear
ramp to the  first Airy ring, while leaving the  central region of the
PSFs  ($0\scnd1$~pixels  in  F110W,  $0\scnd15$~pixels in  F160W,  and
$0\scnd2$~pixels in F205W) unaltered.  The form of the linear ramp was
chosen to be $PSF'(x,y) = S  (y-y_c) PSF(x,y)$, where $S$ is the slope
of the  ramp, and $y_c$ is  the PSF center.   As can be seen  from the
middle-row  panels Figure~\ref{fig1}, the  region above  (towards high
row  number)   component  A  is  free  of   contamination  from  other
components, as is the region below  component B; it was to the data in
these two regions that we fit the multiplicative slope $S$, which gave
$S=200\%/{\rm arcsec}$ in F110W,  $S=120\%/{\rm arcsec}$ in F110W, and
$S=66\%/{\rm arcsec}$ in F205W.

\begin{deluxetable}{lcccc}
\footnotesize
\tablenum{1}
\tablecaption{Photometry and  relative positions of  the components of
\apm.}\label{table1}
\tablewidth{0pt}
\tablehead{
\colhead{quantity} & \colhead{Image} & \colhead{F110W} & 
\colhead{F160W} & \colhead{F205W}}
\startdata
Magnitude                         &A&$13.450 (13.450\pm 0.017)$&$13.091 (13.105\pm 0.011)$&$12.317 (12.240\pm 0.027)$\\
                                  &B&$13.738 (13.755\pm 0.016)$&$13.357 (13.372\pm 0.011)$&$12.603 (12.509\pm 0.011)$\\
                                  &C&$15.368 (15.240\pm 0.026)$&$14.920 (14.743\pm 0.024)$&$14.255 (14.028\pm 0.125)$\\
Color (${\rm M_{F110W}-M}$)       &A&$ 0.                     $&$ 0.359  (0.345\pm 0.020)$&$ 1.133  (1.210\pm 0.032)$\\
                                  &B&$ 0.                     $&$ 0.381  (0.383\pm 0.019)$&$ 1.135  (1.246\pm 0.019)$\\
                                  &C&$ 0.                     $&$ 0.448  (0.497\pm 0.035)$&$ 1.113  (1.212\pm 0.128)$\\
Flux (mJy)                        &A&$ 2.577 (2.579 \pm 0.040)$&$ 3.825 (3.775 \pm 0.033)$&$ 4.035 (4.334 \pm 0.108)$\\
                                  &B&$ 1.977 (1.947 \pm 0.026)$&$ 2.994 (2.952 \pm 0.029)$&$ 3.101 (3.381 \pm 0.032)$\\
                                  &C&$ 0.441 (0.496 \pm 0.012)$&$ 0.710 (0.836 \pm 0.018)$&$ 0.677 (0.840 \pm 0.101)$\\
$\nu F_\nu (10^{-15} {\rm W/m^2})$&A&$ 6.847 (6.850 \pm 0.106)$&$ 7.140 (7.046 \pm 0.062)$&$ 5.844 (6.277 \pm 0.156)$\\
                                  &B&$ 5.252 (5.172 \pm 0.069)$&$ 5.589 (5.511 \pm 0.053)$&$ 4.491 (4.897 \pm 0.047)$\\
                                  &C&$ 1.170 (1.317 \pm 0.031)$&$ 1.325 (1.560 \pm 0.034)$&$ 0.981 (1.217 \pm 0.146)$\\
Relative Flux $F/F_A$             &A&$ 1.                     $&$ 1.                     $&$ 1.                     $\\
                                  &B&$ 0.767 (0.755 \pm 0.015)$&$ 0.783 (0.782 \pm 0.010)$&$ 0.768 (0.780 \pm 0.021)$\\
                                  &C&$ 0.171 (0.192 \pm 0.005)$&$ 0.186 (0.221 \pm 0.005)$&$ 0.168 (0.194 \pm 0.024)$\\
\tableline \\
$x_A-x_B$                         &B&$ 0\scnd220  (0\scnd220\pm 0\scnd001)$&$ 0\scnd220  (0\scnd220\pm 0\scnd001)$&$ 0\scnd214  (0\scnd214\pm 0\scnd001)$\\
$y_A-y_B$                         &B&$ 0\scnd307  (0\scnd308\pm 0\scnd001)$&$ 0\scnd309  (0\scnd309\pm 0\scnd001)$&$ 0\scnd308  (0\scnd309\pm 0\scnd002)$\\
$((x_A-x_B)^2+(y_A-y_B)^2)^{1/2}$
                                  &B&$ 0\scnd378  (0\scnd378\pm 0\scnd001)$&$ 0\scnd379  (0\scnd380\pm 0\scnd001)$&$ 0\scnd375  (0\scnd375\pm 0\scnd001)$\\
$x_A-x_C$                         &C&$ 0\scnd060  (0\scnd065\pm 0\scnd002)$&$ 0\scnd055  (0\scnd056\pm 0\scnd002)$&$ 0\scnd066  (0\scnd055\pm 0\scnd004)$\\
$y_A-y_C$                         &C&$ 0\scnd139  (0\scnd142\pm 0\scnd002)$&$ 0\scnd130  (0\scnd132\pm 0\scnd001)$&$ 0\scnd141  (0\scnd129\pm 0\scnd008)$\\
$((x_A-x_C)^2+(y_A-y_C)^2)^{1/2}$
                                  &C&$ 0\scnd152  (0\scnd156\pm 0\scnd002)$&$ 0\scnd141  (0\scnd143\pm 0\scnd001)$&$ 0\scnd156  (0\scnd140\pm 0\scnd007)$\\
$\alpha_A - \alpha_B$             &B&$ 0\scnd196  (0\scnd197\pm 0\scnd001)$&$ 0\scnd198, (0\scnd198\pm 0\scnd001)$&$ 0\scnd199, (0\scnd200\pm 0\scnd002)$\\
$\delta_A - \delta_B$             &B&$ 0\scnd323  (0\scnd323\pm 0\scnd001)$&$ 0\scnd324, (0\scnd324\pm 0\scnd001)$&$ 0\scnd318, (0\scnd318\pm 0\scnd001)$\\
$\alpha_A - \alpha_C$             &C&$ 0\scnd104  (0\scnd105\pm 0\scnd002)$&$ 0\scnd098, (0\scnd099\pm 0\scnd001)$&$ 0\scnd104, (0\scnd097\pm 0\scnd008)$\\
$\delta_A - \delta_C$             &C&$ 0\scnd110  (0\scnd116\pm 0\scnd002)$&$ 0\scnd102, (0\scnd103\pm 0\scnd002)$&$ 0\scnd116, (0\scnd101\pm 0\scnd005)$\\
\enddata
\tablenotetext{}{The entries  not given  in brackets are  derived from
the  median-combined image stacks,  while values  in brackets  are the
mean  and  RMS scatter  of  measurements  on  individual frames.   The
relative  positions  are  listed  in  the system  of  the  coordinates
$(x,y)$, which are aligned, respectively,  with the CCD row and column
directions.   Also   listed  are  the  positions   in  the  equatorial
coordinates relative to object A.}
\end{deluxetable}

Using these  PSF models  with an improved  estimate of  the brightness
distribution  around   the  first  Airy  ring,   we  obtained  ALLSTAR
measurements of  the magnitudes and positions of  the three components
A, B and  C on the median-combined image stacks  in each color.  These
measurements are listed in Table~1.

We now summarize the observational results.
\begin{itemize}
\item
The reduced chi-squared value of  the PSF fits to the three components
in the median-combined stacked frame  is $\chi^2 < 2$, indicating that
to good  approximation, the three  components are point  sources.  The
residuals of these PSF fits  are displayed in the bottom-row panels of
Figure~\ref{fig1}.
\item
The  colors  of  components  A  and  B are  identical  to  within  the
uncertainties.  If  we accept that  the measurements from  the stacked
frames are more reliable than  the mean value of the measurements from
individual frames, then the colors  of component C are also consistent
with those of A and B.   The simplest hypothesis is therefore that the
components A, B and C have identical colors.
\item
Given that the colors of  the components are identical, we can average
their relative  brightnesses over the  three passbands.  We  find then
that the relative brightness of components A and B is $0.773\pm 0.007$
($0.772\pm 0.012$),  where the  value in brackets  is the mean  of the
measurements on  unstacked frames, and  the uncertainties are  the RMS
scatter  in  the  three  measurements.   The  relative  brightness  of
components A and C is  $0.175\pm 0.008$, ($0.202\pm 0.013$).  Thus, we
find good consistency between the measurement methods, which indicates
that  the relative  brightnesses of  components A,  B and  C  are well
constrained by the data.
\item
The fluxes of the components in $\nu F_\nu$ are approximately constant
as a function of wavelength (slightly brighter in H).
\item
Averaging  over  the  position  measurements  in  each  passband,  the
distance from A  to B is $0.377\pm 0.002$  ($0.379\pm 0.001$), and the
distance  from A  to C  is $0.150\pm  0.006$ ($0.146\pm  0.007$).  The
location of component  C is not on the straight  line connecting A and
B.  This can be seen by measuring the position of component C in a new
coordinate system  $(X,Y)$, where $X$ points  from A to B,  and $Y$ is
orthogonal  to $X$  (and points  in a  north-westerly  direction).  In
these  coordinates, $X_C=0.146\pm 0.007$  ($X_C=0.143\pm  0.007$)  and
$Y_C=0.031\pm  0.001$   ($Y_C=0.030\pm  0.0005$).   While   these  RMS
uncertainties  in the  $Y_C$ positions  likely underestimate  the true
uncertainty, it  is clear that to  a high confidence  level, the three
components of \apm\ are not co-linear.
\end{itemize}

\subsection{VLA}

We  obtained 3.6~cm  radio observations  of APM~08279+5255,  using the
hybrid  BnA~configuration  of  the  NRAO\footnote{The  National  Radio
Astronomy  Observatory   is  a   Facility  of  the   National  Science
Foundation,  operated   under  cooperative  agreement   by  Associated
Universities, Inc.}
Very Large Array (VLA), on 18$^{\rm th}$ June, 1998.  We integrated on
the  target  for  about   2.5~hours,  divided  into  half-hour  scans,
interleaved   with  a  nearby   phase  calibrator   (B0820+560).   The
interferometer  data   were  calibrated  and   mapped  using  standard
AIPS\footnote{   AIPS  (Astronomical   Image  Processing   System)  is
distributed by NRAO.}
procedures, and  the flux densities were calibrated  to 3C\,147 (Baars
et al.  1977).   The target proved to be very faint  in the radio, and
no significant  polarized emission was  detected.  Thus only  a single
iteration of CLEAN mapping was performed, without self-calibration.

The resulting radio map (see Figure~3) shows a very faint source, with
a peak flux  density of only 0.26\,mJy/beam, where  the off-source map
RMS  was 0.013\,mJy/beam, close  to the  expected thermal  noise.  The
peak   of    the   radio   emission    is   at   $\alpha$=08:27:58.00,
$\delta$=+52:55:26.9 (B1950), displaced by $\sim0\farcs6$ from the HST
A-image  position,  probably due  to  a  combination  of HST  and  VLA
astrometric   errors.   The   total  VLA   flux  of   the   source  is
$0.45\pm0.03$\,mJy, integrated over  a $\sim1\arcsec$ square aperture.
There is also a  marginally detected source (peak SNR$\sim5$), roughly
$3\arcsec$   to  the  southwest,   with  a   total  flux   density  of
$0.09\pm0.03$\,mJy.  To  best display  the radio structure,  given the
elliptical natural beam, we  used a $0\farcs4$ FWHM circular restoring
beam.

The radio  source is clearly resolved  along the A--B  image axis.  We
fitted two Gaussian components to a $1\arcsec$ square region enclosing
the radio source, with a relative offset fixed to the HST A--B offset,
and  widths  matched  to  the  $0\farcs4$ convolving  beam.   The  two
components had an  A/B flux density ratio of  0.6, consistent with the
HST  image brightness  ratio of  0.7, but  with a  noticeable residual
between  them.   Adding  a  third  component  at  the  C-image  offset
decreased the fit $\chi^2$ by $\sim30\%$, and yielded B/A and C/A flux
ratios  of   0.9  and  0.5   respectively.   However,  we   could  not
definitively confirm the C-image:  the fit $\chi^2$ remained less than
twice the  best value  provided that the  C image contained  less than
twice the A image flux density.

\section{Gravitational Lensing}\label{lens}
\subsection{Image Configuration}
Using ground-based data,  both Irwin et al.  (1998)  and Ledoux et al.
(1998) determined  that \apm\  comprised a  pair of  point-like images
separated by  $\sim0\scnd4$.  With the  limited information available,
namely the  image separation  and relative brightnesses,  both modeled
the  lensing configuration as  a singular  isothermal sphere  and this
suggested that the quasar source in  \apm\ is magnified by a factor of
20--40.   Taking this  into account,  intrinsically \apm\  still ranks
amongst the brightest systems known.

The data  presented here  greatly enhances our  view of  \apm, clearly
resolving the  system into a  pair of bright point-like  images either
side of a fainter third image.  But before the degree of gravitational
lensing in this system can be  fully explored, the nature of the third
image  must be  investigated.  Two  possibilities  present themselves;
either it is the lensing galaxy which is responsible for splitting the
bright image pair, or it represents a true `third' image of the quasar
source.

Several  lines of argument   point towards the   latter possibility.  First,
morphologically  image C is point-like  and it possesses identical colors to
the  brighter components.  The   spectra of \apm\   with both the 2.5m Isaac
Newton  Telescope  (\cite{ir98})      and  the  10m   Keck   I     Telescope
(\cite{sara99a,sara99}) reveal  the presence for two  \ion{Mg}{2} absorption
systems at z=1.18  and z=1.81.  If we adopt  the lower of these redshifts as
the potential redshift  of the lensing  galaxy, then its distance modulus is
$44.6$ ($H_0=50 {\rm km s^{-1} Mpc^{-1}}$, $q_0 = 0.5$).  Thus if image C is
the lens, its absolute magnitude  would be have to  be prodigious, ${\rm M_J
\sim -32}$ (note that the K-correction is approximately  zero, as the SED of
component C is flat in $\nu f_\nu$).  This is inconsistent with the expected
Faber-Jackson  luminosity of   the  lens given  its small  $\sim   130 \kms$
velocity dispersion (Irwin \etal\   1998), predicted by a simple  isothermal
sphere  lensing  model.  Finally, examination  of  the Keck  I HIRES spectra
(\cite{sara99a,sara99}),      which  have    a  resolution  of   $\sim6\kms$
(0.04\AA/pixel),  reveals a number  of  absorption  systems both along   the
line-of-sight  to the  quasar, the   \lya\ forest,  and associated  with the
quasar source, the  broad  absorption lines.   In  large spectral intervals,
through regions of significant optical depth, \lya\ absorption is saturated,
and the spectrum is effectively black with a  signal that is consistent with
zero.  Since flux from the  (foreground) lensing galaxy  must fill in  these
troughs, their  darkness   can be used  to    place an upper  limit  to  the
brightness of  the lens.  Indeed, between 5500\AA\  to 5900\AA,  the darkest
200~pixel-wide  region is centered  at   5769.2\AA, and the $3\sigma$  upper
limit to  the mean flux in  this 200~pixel region is  600 times fainter than
the  mean flux over  the interval 5500\AA\ to  5900\AA\ (we have adopted the
\cite{sara99a} error spectrum   as a reasonable  estimate of  all sources of
noise  in the  spectrum,  including the uncertainty   arising from scattered
light  in  the HIRES  spectrograph).    This  implies that   the --  as  yet
undetected -- lensing  galaxy must be at  least $\sim 7$~magnitudes  fainter
than the quasar,  that is, ${\rm V  \simgt 22}$.  Thus, the  hypothesis that
image C  is the lensing galaxy leads  to an unrealistically red color: ${\rm
V-K \simgt 8}$.  These  facts provide compelling  evidence that C is a third
image   of the high    redshift quasar source.   This,   however, can not be
conclusively  demonstrated without further observations  of \apm, and in the
following  sections, which deal  with gravitational  lens modeling, both the
third image and   lensing galaxy possibilities   as a source  for  C will be
considered.

\subsection{Modeling}
Given \apm's  apparent position as the most  luminous object currently
known,  it  is   important  to  determine  how  much   the  action  of
gravitational lensing  is enhancing our  view of this  distant source.
The  excellent  resolution  of  the  NICMOS  images  reveals  a  third
point-like  source  between  the  brighter  quasar  image,  and  while
available evidence  points to this being  a third image  of the quasar
source,  the possibility  that  this emission  arises  in the  lensing
galaxy  is not  ruled  out.  In terms  of  modeling the  gravitational
lensing in \apm\ both possibilities are considered.

\subsubsection{C as a Third Quasar Image}\label{gfl_model}
If image C represents a third  image of the quasar source, then extant
observations have  failed to detect  the lensing galaxy and  hence its
characteristics  and position  relative to  the quasar  images remains
unknown to us.  However, due to  the fact that the image positions are
not  co-linear, we know  that this  lens must  possess some  degree of
asymmetry, either due to some intrinsic ellipticity or shearing from a
nearby companion.   Similarly, to produce the  relatively bright third
image, a finite core radius is required~(\cite{wa93}).  We would wish,
therefore,  to   employ  a  finite   core,  and  an   elliptical  mass
distribution  in modeling  the gravitational  lensing in  this system.
Any  such model,  however, requires  a  minimum of  7 parameters  (two
source positions, an ellipticity, core radius, mass orientation, slope
and  normalization), while the  data only  offer 5  constraints (three
relative positions and two relative image brightnesses).  To this end,
we  constructed a  simple model  with  a general  search of  parameter
space.

To reduce  parameter   space,  the mass  distribution   is taken  as   being
isothermal at large radii, turning over in the inner regions to give a core.
Sampling   the  image  characteristics for  a    range of  source positions,
ellipticities    and core radii, resulted     in   the model presented    in
Figure~\ref{fig4}.  The left-hand panel presents  the position of the source
relative to  the caustic distribution  in the source plane.   The right-hand
panel presents the image plane  with the corresponding  critical lines.  The
small  circles represent  the resultant positions  of  the images; these are
superimposed on  the NICMOS image.   Both the  relative image  positions and
magnitudes match those outlined in  Table~\ref{table1}.  The lensing mass is
oriented 76$^{\rm o}$ east  of north, with a  small deviation from  circular
symmetry $(\epsilon\sim0.01)$. The    core  radius  of the   lens   is $\sim
0\scnd21$,  and assuming that  this galaxy is at a  redshift  of z=1.18, its
mass interior to   the Einstein ring  is  ${\rm  \sim 2\times10^{10}\msun}$.
While this recovers  the observed   image plane  characteristics it  is,  of
course, highly degenerate, but it does demonstrate  that \apm\ is consistent
with being a gravitational lensed system.

We do  note, however, that more  complex lens models  are possible and
\apm\  may   represent  lensing  by  a   `naked  cusp'  (\cite{wa93}).
Similarly, the location of at least one other lensing system along the
line-of-sight to \apm\  can lead to a more  complex caustic network in
the  source   plane.   With  either  of  these,   the  observed  image
configuration  can  be  reproduced  with quite  different  degrees  of
magnification.    The  true  nature   of  the   gravitational  lensing
configuration in  this system will remain uncertain  until the lensing
galaxy has been detected.

\subsubsection{C as the Lensing Galaxy}\label{joes_model}
The  observed   three  image   configuration  is  consistent   with  a
fundamental theorem  of gravitational  lensing which dictates  that an
odd  number of  images are  {\it always}  formed by  non-singular mass
distributions   (\cite{bu81}).    It  is,   however,   at  odds   with
observations of most gravitationally lensed systems, as these are seen
to    possess    an    even    number   of    images    (c.f.     {\tt
http://cfa-www.harvard.edu/castles/});  this seemingly incompatibility
between gravitational  lens theory and observations  is usually solved
by invoking  a (near-)singular  core in the  lensing galaxy  which can
drive the  magnification of one  of the images to  zero (\cite{na86}).
The inverse of  this suggests that the lensing  mass distribution must
contain a  finite core radius if the  image at C is  to be appreciably
magnified;  this is  exactly  the situation  with  the previous  model
(Section~\ref{gfl_model}).

Unless this  is the  first true bona-fide  three image  lensed system,
however, we must also consider  the case that C represents the lensing
galaxy and  the `true' third image  has been demagnified  by an almost
singular core.  With  this, the lensing potential of  C was modeled as
an elliptical,  isothermal potential (c.f.~\cite{ko89}).   The lensing
configuration  offers 5  constraints: two  positions, relative  to the
lens, for each image, and  the relative image brightnesses.  The model
has 3 free parameters,  the ellipticity, orientation and normalization
of the mass  distribution. Using C as the position of  the lens is not
used as  a constraint,  rather it is  also considered another  pair of
free parameters. With 5 free parameters and 5 constraints a search for
an exact model fit can be made.

The resulting minimized model  lensing configuration for this model is
presented in Figure~\ref{fig5}; here  the grey represents the position
of the  source relative  to the lensing  caustics, while  the observed
images and critical  lines are solid.  The best  position for the lens
is  $0\scnd0989$ West and  $0\scnd1026$ South  of image  A, consistent
with the position of image C (Table~\ref{table1}).  The mass has a 1-D
velocity dispersion  of 126$\kms$ and is oriented  at 104$^\circ$ east
of  north,  an  offset  of  $28^\circ$ from  the  model  presented  in
Section~\ref{gfl_model}, although the  ellipticity of the mass profile
is substantial, with $\epsilon=0.2$; the differences in the models are
understandable given the different interpretations of the nature of C.
The  source  has  an  impact  parameter,  relative  to  the  lens,  of
$0\scnd01$ (SSW).   Again, the  location of the  source is  similar to
that seen in Section~\ref{gfl_model}.  This analysis illustrates, from
a  gravitational lensing  point of  view, that  C could  represent the
lensing galaxy.

But what  of the nature  of C? As  discussed previously, if it  is the
lensing system then its brightness indicates that it is not a `normal'
galactic  system.  One  possibility is  that  C is  itself a  luminous
quasar,  explaining  both   its  apparent  brightness  and  point-like
appearance.    Such   quasar-quasar   lensing   has   been   addressed
(\cite{wa97}), although only  high resolution imaging and spectroscopy
will uncover C's true nature.

\subsection{The Intrinsic Luminosity of \apm}\label{lumo}
Using the models described  above, the degree to which  the quasar source is
magnified    can be   determined.    Considering    the   first  model    of
Section~\ref{gfl_model}, which  assumes  image C  is a  third  image of  the
quasar,  the    total  magnification  of the  quasar    source is  $\sim90$.
Similarly, the model described in Section~\ref{joes_model}, in which image C
represents  the lensing galaxy responsible   for splitting the quasar  light
into the  A and  B  images, the  corresponding  magnification of the  quasar
continuum source is  $\sim7.5$. We expect the   magnification of the  far-IR
continuum source to be  a factor of 2--3 times  lower than the magnification
of the  quasar  continuum source, due to  the  larger size of  the  emission
region.   Taking  these into account, the  intrinsic  luminosity of \apm\ is
$\sim10^{14}\rightarrow{10^{15}\lsun}$ and it  retains its place amongst the
most luminous systems currently known.

\section{Conclusions}\label{conclusions}
We  have presented new  observations of  the ultraluminous  BAL quasar
\apm, using both  NICMOS on the Hubble Space Telescope  and the VLA at
3.5cm. These  clearly demonstrate the composite nature  of the system,
separating  \apm\  into a  pair  of images,  A  and  B, of  comparable
brightness.  The  NICMOS images  also reveal the  presence of  a third
point-source,  C,  between the  brighter  two.   The  VLA image  shows
structure corresponding  to the background quasar being  a faint radio
source  and we  conclude that  intrinsically  \apm\ is  a radio  quiet
quasar with  ordinary radio properties.  Although the  radio source is
resolved  along A--B,  we  can  not determine  from  these radio  data
whether it comprises the A+B images or the A+B+C images.

There are two possible interpretations of the source of C: either this
central image is  a detection of the foreground  lensing galaxy, or it
represents  a  third  image  of  the high  redshift  quasar.   Several
arguments  indicate that  this  latter proposition  is  a more  likely
description of the available data.  A lensing model in which the three
observed images  are true images  of the quasar source  was presented.
This  recovers   both  the   image  configuration  and   and  relative
brightnesses with a total lensing magnification of $\sim90$.

An alternative model in which image  C is the lensing galaxy, was also
explored.   The position  of C  was found  to be  coincident  with the
position of the lensing galaxy  predicted from a simple lensing model,
indicating  that this scenario  may not  be implausible.   However, to
account for its  apparent brightness, C must also  be an intrinsically
luminous source.  If this latter  model is correct, the simple lensing
configuration it  offers, with a  corresponding time delay  of 5~days,
provides an ideal  example of a ``golden lens''  (Williams \& Schecter
1997) with  which  ${\rm  H_o}$  can be  determined  from  photometric
monitoring.

Considering  the  magnification of  the  above  models, the  intrinsic
luminosity  of \apm\ is  $\sim10^{14}\rightarrow10^{15}\lsun$, placing
it amongst the most luminous systems currently known, and its apparent
brightness  makes it  ideal  to  study aspects  of  the high  redshift
Universe (\cite{hi99,sara99,sara99a}).

Recent near and mid-IR KECK  images were brought to our attention just
prior  to submitting  this article  (Egami  et al.   1999).  Taken  in
excellent seeing, these also clearly  reveal the presence of the third
image in the same location and with a similar brightness to that found
in the  the NICMOS images.   Given its apparent brightness,  they also
conclude that this represents a third image of the quasar, rather than
the lensing galaxy.  They also  construct a very similar lensing model
to the one presented in this paper, although ours possesses a slightly
higher total  magnification (90 compared  to their 71).   Overall, our
conclusions are in excellent agreement with theirs.

\acknowledgements  We  thank the  anonymous  referee for  constructive
comments,  and  Dr.  E.  Egami  for  pointing  out  an  error  in  the
orientation of one of our lensing models.

\newpage

\begin{figure*}
\centerline{
}
\caption[]{NICMOS  color-composite  image of  the  \apm\ system.   The
orientation and scale is the  same as Figure~\ref{fig1}.  Again, C can
be clearly seen between the brighter images.  }
\label{fig2}
\end{figure*}

\begin{figure*}
\centerline{
\psfig{figure=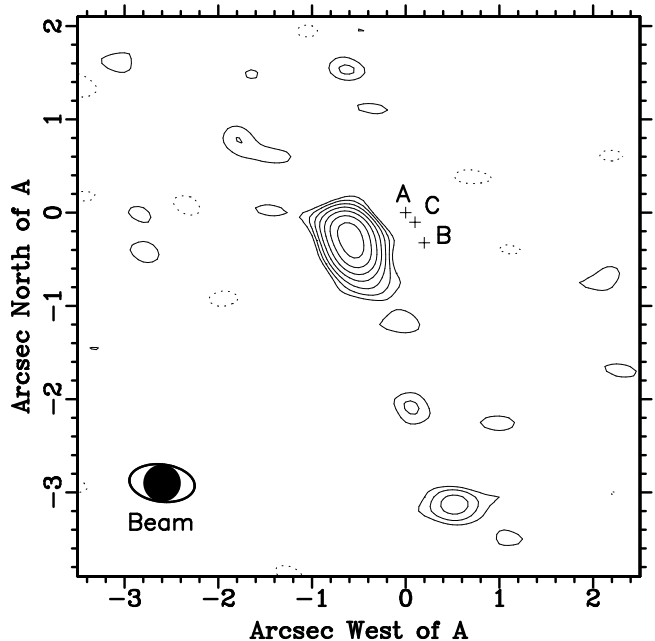,bbllx=34pt,bblly=132pt,bburx=530pt,bbury=630pt,clip=yes,width=3.5in}
}
\caption[]{  VLA map  of APM08279+5255  at 3.6cm,  obtained  using the
hybrid  BnA  configuration.   The  contours  increase  by  factors  of
$\sqrt{2}$   from   twice   the   off-source  RMS   noise   level   of
0.013\,mJy/beam.  The  HST image  components are marked  with crosses,
and    all    position    offsets    are    relative    to    A,    at
$\alpha$=08:31:41.64,$\delta$=+52:45:17.5\,.   The   radio  source  is
offset by  $\sim0.6\arcsec$, probably due  to VLA and  HST astrometric
errors.   The  natural  and  convolving  beams  (FWHM)  are  shown  at
bottom-left, respectively as open and filled ellipses.  }
\label{fig3}
\end{figure*}

\begin{figure*}
\centerline{
}
\caption[]{ The  left-hand panel presents source plane  with the image
position relative to the caustic structure of the lens model discussed
in   Section~\ref{lens}.    The    right-hand   panel   presents   the
corresponding critical  lines over the  image plane, coupled  with the
image  positions,  overlaid  on   the  NICMOS  images  of  \apm.   The
orientation of both frames  is identical to that of Figures~\ref{fig1}
and~\ref{fig2}.}
\label{fig4}
\end{figure*}

\begin{figure*}
\centerline{
\psfig{figure=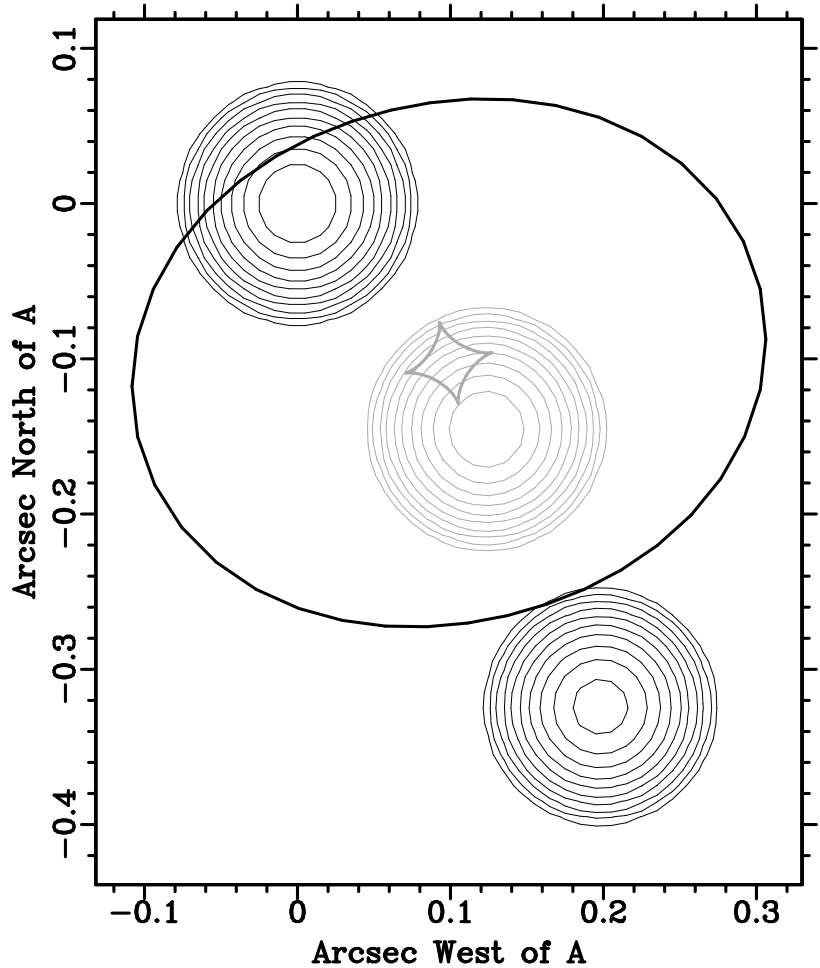,width=5in,angle=0}
}
\caption[]{ If  C is the  lensing galaxy, a  simple lens model  can be
constructed to  account for the observed image  configuration; this is
illustrated  above.   Here,  the  light grey  represents  the  caustic
structure  and source position,  while the  solid lines  represent the
observed   image   configuration  and   critical   line.   The   total
magnification  of a  point-like source  with this  model is  only 7.5,
substantially less than  the 90 expected if C is  actually an image of
the quasar.  In this  plot, North is to the top, while  East is to the
left and  the coordinate system is  centered on image  A.  The source
and images are point-like but have been smoothed with a $0\scnd03$ 
Gaussian to represent the NICMOS resolution.}
\label{fig5}
\end{figure*}


\newpage

\begin{thebibliography}{}
%
\bibitem[Barvainis et al.\ 1995]{ba95}
  Barvainis, R., Antonucci, R., Hurt, T., Coleman, P. \& Reuter, H.-P.,
  1995, \apj\, 451, 9
%
\bibitem[Broadhurst \& Leh\'{a}r 1995]{br95}
  Broadhurst, T. \& Leh\'{a}r, J., 
  1995, \apj\, 450, 41
%
\bibitem[Burke 1981]{bu81}
  Burke, W.L.,
  1981, \apj\, 244, L1
%
\bibitem[Downes et al.\ 1998]{do98}
  Downes, D., Neri, R., Wikland, T., Wilner, D.J. \& Shaver, P.,
  1998, \apj\, 513, L1
%
\bibitem[Egami et al.\ 1999]{eg99}
 Egami, E., Neugebauer, G., Soifer, B. T., Matthews, K., Ressler, M. \&
 Becklin, E. E.,
 1999, \apj\, Submitted
%
\bibitem[Eisenhardt et al.\ 1996]{ei96}
  Eisenhardt, P. R., Armus, L., Hogg, D. W., Soifer, B. T.,
  Neugebauer, G. \& Werner, M. W.,
  1996, \apj\, 461, 72
%
\bibitem[Ellison et al. 1999a]{sara99a}
        Ellison, S. L., Lewis, G. F., Pettini, M., Sargent, W. L. W.,
        Chaffee, F. H., Foltz, C. B., Rauch, M. \& Irwin, M. J.
        1999a, \pasp, {\it In Press}
%
\bibitem[Ellison et al. 1999]{sara99}
        Ellison, S. L., Lewis, G. F., Pettini, M.,
        Chaffee, F. H. \& Irwin, M. J.
        1999, \apj, 520, 456
%
\bibitem[Hines et al. 1999]{hi99}
        Hines, D. C., Schmidt, G. D. \& Smith, P. S.,
        1999, \apj, 514, L91
%
\bibitem[Irwin et al.\ 1998]{ir98}
  Irwin, M. J., Ibata, R. A., Lewis, G. F. \& Totten, E. J.,
  1998, \apj\, 505, 529
%
\bibitem[Kneib et al.\ 1998]{kn98}
  Kneib, J.-P., Alloin, D., Mellier, Y., Guilloteau, S., 
  Barvanis, R. \& Antonucci, R.,
  1998, \aap\ 329, 827
%
\bibitem[Kochanek, Blandford, Lawrence \& Narayan 1989]{ko89} 
	Kochanek, C. S., Blandford, R. D., Lawrence, C. R. \& Narayan, R.,
	1989,\mnras, 238, 43 
%
\bibitem[Ledoux et al.\ 1998]{ld98}
  Ledoux, C., Theodore, B., Petitjean, P., Bremer, M.N.,
  Lewis, G.F., Ibata, R.A., Irwin, M.J. \& Totten, E.J.,	
  1998, \aap\, 339, L77
%
\bibitem[Lewis et al.\ 1998]{le98}
  Lewis, G.F., Chapman, S.C., Ibata, R.A., Irwin, M.J. \& Totten, E.J.,	
  1998, \apj\, 505, L1
%
\bibitem[Magain et al.\ 1988]{ma88}
  Magain, P., Surdej, J., Swings, J.-P., Borgeest, U. \& Kayser, R.,
  1988, \nat\, 334, 325
%
\bibitem[Narasimha Subramanian \& Chitre 1986]{na86}
  Narasimha, D., Subramanian, K. \& Chitre, S. M.,
  1986, \nat, 321, 45 
%
\bibitem[Rowan-Robinson et al.\ 1991]{ro91}
  Rowan-Robinson, M., Broadhurst, T., Oliver, S.J., Taylor, A.N.,
  Lawrence, A., McMahon, R.G., Lonsdale, C.J., Hacking, P.B. \&
  Conrow, T., 1991, \nat\, 351, 719
%
%
\bibitem[Seitz et al.\ 1998]{se98}
  Seitz, S., Saglia, R. P., Bender, R., Hopp, U., Belloni, P. \& Ziegler, B.,
  1998, \mnras\, 298, 945
%
\bibitem[Stetson 1987]{st87} 
Stetson, P. B., 
1987, \pasp, 99, 191 
%
\bibitem[Wallington \& Narayan 1993]{wa93}
  Wallington, S. \& Narayan, R.,
  1993, \apj\, 403, 517	
%
\bibitem[Wampler 1997]{wa97} 
	Wampler, E. J.,
 	1997, \apjl, 476, L55 
%
\bibitem[Williams \& Lewis 1996]{wi96}
  Williams, L. L. R. \& Lewis, G. F., 
  1996, \mnras\, 381, L35  
%
\bibitem[Williams \& Schecter 1997]{wi97}
  Williams, L. L. R. \& Schecter, P. L., 
  1997, {\it Astronomy \& Geophysics}, 38, 10
%
\bibitem[Yee et al.\ 1996]{ye96}
  Yee, H. K. C., Ellingson, E., Bechtold, J., 
  Carlberg, R. G. \& Cuillandre, J.-C., 
  1996, \aj\, 111, 1883 
%
\end{thebibliography}
\end{document}